\documentclass[12pt,a4paper]{article}
\usepackage{amssymb}
\usepackage{graphicx}
\usepackage{amsmath}

\newtheorem{theorem}{Theorem}
\newtheorem{acknowledgement}[theorem]{Acknowledgement}

\input{tcilatex}

\begin{document}

\title{INFLATIONARY LAMBDA-UNIVERSE WITH TIME-VARYING FUNDAMENTAL \ CONSTANTS%
}
\author{Marcelo Samuel Berman$^{1}$ and Luis A. Trevisan$^{2}$. \\
$^{1}$ Instituto Albert Einstein/Latinamerica\\
Av. Candido Hartman, 575 \#17 \\
80730-440 \ \ Curitiba / PR \ \ \ Brazil\\
email: msberman@institutoalberteinstein.org\\
\ \ \ \ \ \ \ \ \ \ marsambe@yahoo.com\\
$^{2}$Universidade Estadual de Ponta Grossa (UEPG)\\
DEMAT-CEP 84010-330-Ponta Grossa-PR-Brazil\\
email: latrevis@uepg.br}
\maketitle

\begin{abstract}
We solve the Universe for dark energy represented by a variable cosmological
\ constant,  along with a JBD (Jordan-Brans-Dicke) model with tine varying
speed of light, entailing variable fine structure \ constant.  Inflation is
taken as the prevailing scenario, but we provide food for thought, to
discuss how this model with fundamental \ constants  applies to the present
accelerating phase. Along with a curious discussion of a possible Planck%
\'{}%
s time not coincident with ten to the power -43, in seconds, this model is
full of novelties.

\end{abstract}

\begin{center}
\bigskip \newpage

{\LARGE INFLATIONARY LAMBDA-UNIVERSE }

\medskip

{\LARGE WITH TIME-VARYING}

\medskip

{\LARGE \ FUNDAMENTAL CONSTANTS}

\medskip

{\large BY MARCELO S. BERMAN AND LUIS A. TREVISAN}
\end{center}

{\large \bigskip }

Dark energy is possibly represented by a time-varying cosmological
"constant". On the other hand, the fact that the equation of state of the
present Universe is very near $p=-\rho $ , where $p,\rho $ \ are cosmic
pressure and energy density, makes us think that inflationary scenario with
exponential scale factor could not only be of importance as early Universe
phase, but could also be important for the present accelerating Universe,
because the deceleration parameter would be negative, as required by modern
developments in the observational front. The present paper will deal with
such questions. We extend here the results of J.D.Barrow , who in a series
of papers authored by him alone, or \ with collaborators, dealt with
variable \textquotedblleft constants\textquotedblleft\ , cosmological
models, including fine structure "constant" , and inflationary scenario
cosmologies \cite{15}; we calculate the possibility of different than the
accepted values for Planck%
\'{}%
s Universe quantities (like Planck`s time or Planck's energy, etc), due to
the effects of inflation on such time variations.

Weinberg \cite{13}, has reviewed modern cosmology, and we refer to him for
updated information. Riess et al \ \cite{R} found evidence for an
accelerating Universe with an observed deceleration parameter average near $%
-1$ \ , by means of Supernovae observations. Not only were placed
constraints on the Hubble%
\'{}%
s constant but also on the mass density, the cosmological constant (i.e, the
vacuum energy density), the dynamical \ age of the Universe, and most
important in our opinion, the deceleration parameter. In the abstract of one
of the papers published by Riess and his group, there is a comment that
eternally \ expanding models with positive cosmological constants are
favoured unanimously. In the present paper, we shall find a \ theoretical
model of the Universe that could eventually point to such an eternal
Universe with infinite age ( \ --$\infty <t<\infty $ \ \ ), or else it could
be understood as defining an exponential inflationary phase for the early
Universe, when according to our calculations the fine structure constant
should have a huge value, when compared with its present value ( $\ \ \alpha
_{0}\approx 1/137$ $\ ).$ Both possibilities render useful \ the study of \
\ \ \ $q=-1$ \ cosmologies.

On the other \ hand, Webb et al \cite{W} found evidence, in\ the spectra of
distant quasars, for a time varying fine structure constant, spanning 23\%
to 87\% of the age of the Universe, if we suppose that there was a big bang,
say about 18 billion years ago. Webb et al. found that \ \ $\alpha $ \ \ \
was smaller in the past than it is today. This could be explained by time
variations of at least one of the constants that appear in its definition(
the electron charge, the speed of light or Planck%
\'{}%
s constant). They also comment that a common property of unified theories,
when applied to cosmology, is that they predict time-space dependence of the
coupling constants.

We shall show how to accommodate both results into a JBD
(Jordan-Brans-Dicke) framework, where the speed of light is also variable,
as in Barrow \cite{B}. John D. Barrow \cite{15} analyzed generalizations of
General Relativity that incorporates a cosmic time variation of the velocity
of light in vacuum and the Newtonian gravitational constant G, as proposed
by Albrecht and Magueijo \cite{AM}. He found exact solutions for Friedmann
Universes and determined the rate of variation of $c$ required to solve the
flatness and classical cosmological constant problems. Potential problems
with this approach were discussed by Barrow, and Jordan-Brans-Dicke
solutions were also presented.

Barrow \cite{1} has pointed out the possible relevance of scalar-tensor
gravity theories in the study of the inflationary phase during the early
Universe. He obtained exact solutions for homogeneous and isotropic
cosmologies in vacuum and radiation cases, for a variable \ coupling
"constant" , $\ \ \omega =\omega (\phi ),$ \ \ where \ \ $\phi $ \ \ \
stands for the scalar field. Although classical standard Big Bang scenario
imposes entropy conservation, in 1981 A. Guth \cite{G} proposed that the
flatness along with the monopole and the horizon problems could be made to
disappear if the Universe traversed an epoch with negative pressure and
terminated it in a huge entropy increase (old inflation).Many new models
were afterward been invented, in particular the new inflationary, Linde%
\'{}%
s chaotic version of it \cite{2}and extended inflation\cite{15}. Berman and
Som \cite{BerSOM} further have shown that in Brans-Dicke theory \cite{BD}we
can accommodate an inflationary exponential phase with positive pressure.
For accounts on inflation, see, for instance, Linde's \ book\cite{2} or the
updated Weinberg [14].

We define the deceleration parameter, in terms of the scale factor and its
time derivatives as:

\begin{equation}
q=-\frac{\ddot{R}R}{\dot{R}^{2}}.
\end{equation}
It is evident that an exponential scale factor, like

\begin{equation}
R=R_{0}e^{Ht}
\end{equation}
($R_{0}$ and $H$ constants) yields the desired result, i.e.

\begin{equation}
q=-1.
\end{equation}

Before introducing the cosmological "constant", we first work with Barrow%
\'{}%
s equations , which do not contain a $\Lambda $ term and zero tricurvature:

\begin{equation}
H^{2}=\frac{8\pi \rho }{\phi }-\frac{\dot{\phi}}{\phi }H+\frac{\omega }{6}%
\frac{\dot{\phi}^{2}}{\phi ^{2}}
\end{equation}

\begin{equation}
\dot{\rho}+3H\left( \rho +\frac{p}{c^{2}}\right) =0
\end{equation}

\begin{equation}
\ddot{\phi}+3H\dot{\phi}=\frac{8\pi }{(3+2\omega )}\left( \rho -\frac{3p}{%
c^{2}(t)}\right)
\end{equation}%
where \ $\rho ,$ $\ \ p,$ $\ \ \ \phi ,$ $\ \ \omega $ \ \ \ and $\ \ \ k$ \
\ \ stand respectively for energy density, cosmic pressure, scalar field,
coupling constant, and tricurvature. This applies to a Robertson-Walker%
\'{}%
s metric. Equation (5) represents the energy-momentum tensor conservation,
for the original matter fluid.

\bigskip

We find the following solution:

\begin{equation}
\phi =\phi _{0}e^{\beta t}
\end{equation}

\begin{equation}
\rho =A\phi _{0}e^{\beta t}
\end{equation}

\begin{equation}
\beta =-3H\left( 1+\gamma \right)
\end{equation}

\begin{equation}
c=c_{0}e^{\delta Ht}
\end{equation}%
subject to the following conditions :

\begin{equation}
\frac{p}{c^{2}}=\gamma \rho
\end{equation}

\begin{equation}
H^{2}=-\frac{8\pi A}{\left[ 2+3\gamma +\frac{3}{2}\omega (1+\gamma )^{2}%
\right] }
\end{equation}

\begin{equation}
A=\frac{(3+2\omega )(\beta +3H)\beta }{(1-3\gamma )}.
\end{equation}

We remark that the last three conditions were obtained by plugging our
solution into the three field equations obtained by Barrow \cite{B}. \ It is
better to restrict our study to the case $k=0$, in order that we need not
bother with possible $\dot{c}$ terms in the above field equations.

\bigskip If we introduce a cosmological term $\Lambda =\Lambda (t),$ we
resort to Barrow-Bertolami theorem given by Berman\cite{Berman} \cite%
{Berman2007a} which assumes the form of replacement:

\begin{equation*}
\rho \rightarrow \rho +\Lambda /(8\pi )
\end{equation*}%
,

\begin{equation*}
p/c^{2}\rightarrow p/c^{2}-\Lambda /(8\pi )+\dot{\Lambda}\phi /(12\pi \dot{%
\phi})
\end{equation*}%
.In such case, we find a reasonable solution, namely:

\begin{equation*}
\Lambda =\Lambda _{0}e^{\beta t}
\end{equation*}

where \ \ \ $\Lambda _{0}$ \ $=$ constant. If we plug the solution back, we
find that, instead of conditions (12) and (13) we must make the following
replacement:

\bigskip

\begin{equation*}
A\rightarrow A+\Lambda _{0}/(8\pi )
\end{equation*}%
and we must have,

\begin{equation*}
c_{0}^{2}=-3\phi _{0}\gamma
\end{equation*}%
. Likewise, as $c_{0}^{2}>0$ , and , for normal situations, $\phi _{0}>0$
(so that, gravitation is attractive, i.e. , $G>0$ ) we must have $\gamma <0$%
. The weak energy condition is represented by a positive energy density, so
that we must have a negative cosmic pressure. This is tantamount \ to what
is admitted by the most recent observations of the present Universe.

\bigskip

\ On observational grounds, we impose that, for the present Universe

\begin{equation}
p=\epsilon \rho
\end{equation}%
, where $\epsilon \leq -0.9$. The speed of light must be adjusted in order
that for the present Universe, we get the result $300,000$ km/s . \ From
relation (8) , the apparent desire for a decreasing energy density, makes us
impose that $\beta <0$ . From (9) , we would need in such case, to impose
that ,

\begin{equation}
1+\gamma >0
\end{equation}
\ 

Because we define the fine structure "constant" \ as

\begin{equation}
\alpha \equiv \frac{e^{2}}{\frac{h}{2\pi }c}
\end{equation}
we find that

\begin{equation}
\frac{\dot{\alpha}}{\alpha }=-H\delta
\end{equation}%
where of course, \ $H$ \ \ \ stand for \ Hubble `s \ constant for this
model. This means that \ \ $\alpha $ \ \ \ was exponentially larger in the
early Universe than it is today, if $\delta >0$. Good discussions on fine
strucuture time-varying "constant", either through $\dot{c}\neq 0$ or
through $\dot{\epsilon}_{0}\neq 0$ can be found in Berman \cite{Berman2009} 
\cite{Berman2010}.

Planck%
\'{}%
s quantities for energy \ length , mass and time could be drastically
changed for the very early Universe ( Grand Unification Epoch) as a
consequence of \ the time variations of \ $\ \ G$ \ \ and \ $c$. \ \ Take
for example Planck%
\'{}%
s length:

\begin{equation}
\lambda =\left( \frac{hG}{2\pi c^{3}}\right) ^{1/2}
\end{equation}%
We find that, with $G\varpropto \phi ^{-1}$,

\begin{equation}
\lambda \varpropto e^{-(1/2)\left[ \beta +3H\right] t}
\end{equation}
On the other hand, Planck%
\'{}%
s time is \ $\lambda /c$ \ \ so that it would vary like:

\begin{equation}
t_{p}\propto e^{(3/2)H\left[ 1+\gamma -\delta \right] t}
\end{equation}%
This means that with most probability \ \ $t_{p}=$10$^{-43}$ \ \ seconds is
no longer a valid result. For the particular case $\delta =1+\gamma $ , we
recover a constant Planck%
\'{}%
s time. The condition $\delta >0$ means that $\gamma >-1$ , for constant $%
t_{P}$ .

Even with exponential equations for $\rho $ and $p$ , we could adjust
constants in order that an equation of state be obtained \ such that $p=%
\frac{1}{3}\rho $ , in case we would like to include a radiation phase,
during the evolution of the Universe.

As the reader can check, many questions can be posed over this mathematical
and theoretical model. . May be we are just in face of an eternal Universe
model, of the type studied in the golden days of stationary state models. If
this is not the case, we have nevertheless found a model that deserves
attention, and that can point to very interesting developments in the study
of the very early Universe Physics. Finally, our model has zero-pressure,
zero-density and zero-lambda when \ \ $t\rightarrow \infty .$ This model
deserves the qualification of present Universe validity, but also has to do
with the very early Universe inflation, but also applies with minor
modification, to a radiation phase. The novelties in this paper, deserve
attention.

\bigskip

\begin{acknowledgement}
MSB thanks the incentive by Geni, Albert and Paula.
\end{acknowledgement}

\bigskip

Hinshaw , G. et al (2008) Astrophysical Journal Supplement Series, to be
published, ArXiv 0803.0732. See also ArXiv :0812.2720 and ArXiv 0902.3702

\end{document}